# Intrinsic operator time of stochastic systems

Walter Gessner

Unterer Katzenbergweg 7

D-97084 Würzburg

## Abstract

Stochastic systems consisting of a very large number of independent elementary processes of the same kind, especially the radioactive decay, are considered as quantum clocks. By adapting the framework of the previously introduced concept of ideal quantum clocks, the time operator for these systems is derived and discussed.

It is shown that the standard deviation of time measurement by such a stochastic device is bounded from below by the limitation of the number of elementary processes from physical reasons and by the Planck-time. As a result, any time dilatation, whether caused by extreme speed of the quantum clock or by gravitational fields, increases the standard deviation. This reduces the accuracy of time measurements especially in space navigation. In the vicinity of the Schwarzschild spherical shell of a black hole, time measurements are completely blurred and thus impossible.

## 1. Introduction

The concept [1] of an "ideal quantum clock" (IQC) uses an operational definition of time where time arises from the dynamics of physics. Accordingly, each IQC C is a physical process, governed by a Hamiltonian $H_C$ and the corresponding "Schrödinger curve" $\varphi_C(t) = \exp(-itH_C)\varphi_C(0)$. It is assumed that an infinite set of equidistant time points $\tau^n := n\tau$, $n \in \mathbb{Z}$, exists so that the $\varphi_C(\tau^n)$ are pairwise orthonormal in their state space $s_C$ whose basic elements are those $\varphi_C(\tau^n)$. The space $s_C$ itself is no Hilbert space but dense in a Hilbert space. The time operator $T_C$ of the IQC-concept uses the nonvanishing components of $\varphi_C(t)$ with respect to the $\varphi_C(\tau^n)$. $T_C$ proves to be canonically conjugated to $H_C$ so that the time-energy relation holds. The clicks of the IQC are the $\tau^n$. While an IQC uses equidistant time points $\tau^n$ in a strictly periodic system, a series of random events of a stochastic system seems to be completely opposite to the IQC-concept at the first glance.

A stochastic system consists of a very large number M of elementary processes of the same kind. Each individual elementary process randomly falls from a state A into a state B. The probability that any process, which is at the time $t_0$ in A falls within the time interval dt about a time point t with $t > t_0$ into B, is given by

$$p(t)dt := \lambda \exp[-\lambda(t - t_0)]dt \quad \text{with} \quad \int_{t0}^{\infty} p(t)dt = 1, \tag{01}$$

where $\lambda > 0$ describes the vehemence of the decay. The time points $t^n$ of the decays are not equidistant. The expectation value of the distance of two neighboring time points $t^n$ is given by

$$\tau := \tau_0/M, \text{ with } \tau_0 := 1/\lambda. \tag{02}$$

The equidistant time points $\tau^n := n\tau$, with $-N \leq n \leq +N$, $M = 2N + 1$, and the intervals $J^n := [\tau^n - \tau/2; \tau^n + \tau/2]$ are now being introduced.

It is assumed, that exactly one decay, characterized by $t^n$, lies in each of these intervals. Instead of using the decay time points $t^n$, whose distribution corresponds to a Poisson statistic, the intervals $J^n$ are used as the clicks of the quantum clock.

The use of the $J^n$ together with average values of certain physical quantities over these intervals will allow the adaption of the IQC-framework also to stochastic systems and thus the analysis of stochastic devices as quantum clocks.

## 2. Stochastic systems

Summarizing, one has now $\tau := \tau_0/M$, with $\tau_0 := 1/\lambda$, the equidistant time points $\tau^n :=  n\tau$, with $-N \leq n \leq +N$, $M = 2N + 1$, and the intervals $J^n$.

The probability (01) that any process, which is at the time 0 in A falls within the time interval $\tau$ about $\tau^n$ with $\tau^n > 0$ into B, is given by

$$\lambda\tau \exp(-\lambda\tau^n) = \frac{1}{M}\exp(\frac{-n}{M}). \tag{03}$$

According to the IQC-concept it is now assumed that the whole system has a Schrödinger curve $\varphi_C(t)$ in a suitable state space, which regulates the time evolution despite the stochastic character of the processes. These $\varphi_C(t)$, which can at most be specified for particularly simple systems, are not used anywhere concretely, but serve only to identify certain terms, especially transition probabilities.

The transition probability of the whole system from $t = 0$ to $\tau^n > 0$ is now given by $|<\varphi_C(0)|\ \varphi_C(\tau^n)>|^2$. The corresponding transition probability $\frac{1}{M}\exp(\frac{-n}{M})$ describes the transition from $t = 0$ into the interval $J^n$, i.e. an average value. Therefore it suggests

itself to replace $|<\varphi_C(0)|\,\varphi_C(\tau^n)|^2$ by its average value over $J^n$. At first it is therefore assumed that

$$\frac{1}{M}\exp(\frac{-n}{M}) = \tau^{-1}\int\limits_{-\tau/2}^{+\tau/2}du|<\varphi_C(0)|\varphi_C(\tau^n+u)>|^2.$$

Prerequisite so far was $\tau^n > 0$. However, the following applies

$$\int\limits_{-\tau/2}^{+\tau/2}du|<\varphi_C(0)|\varphi_C(\tau^n+u)>|^2 = \int\limits_{-\tau/2}^{+\tau/2}dv|<\varphi_C(0)|\varphi_C(-\tau^n+v)>|^2, \qquad (04)$$

so that now the sign of $\tau^n$ can be set without restriction ($-N \leq n \leq +N$).

Summarizing it is decided:

$$\tau^{-1}\int\limits_{-\tau/2}^{+\tau/2}du|<\varphi_C(0)|\varphi_C(\tau^n+u)>|^2 = \frac{1}{2M}\exp(\frac{-|n|}{M}) \text{ with} \qquad (05)$$

$$\frac{1}{2M}\sum_{n=-\infty}^{+\infty}\exp(\frac{-|n|}{M}) = 1. \qquad (06)$$

The factor ½ in (05) guarantees the norm (06). Because of the very large M it can be summed from $-\infty$ to $+\infty$.

## 3. Transition to the IQC-concept

In the IQC-concept the C-characteristic functions $c^{kn}(u)$ play a decisive role. They are defined by

$$c^{nk}(u) = <\varphi_C(\tau^n)|\varphi_C(\tau^k+u)>, \, u \in [-\tau/2; +\tau/2]. \tag{07}$$

As a result of (05) and especially for t = 0 we assume in summary that

$$\tau^{-1} \int_{-\tau/2}^{+\tau/2} du \, |c^{0n}(u)|^2 = \tau^{-1} \int_{-\tau/2}^{+\tau/2} du \, |<\varphi_C(0)|\varphi_C(\tau^n+u)>|^2 = \frac{1}{2M} \exp(\frac{-|n|}{M}), \tag{08}$$

where -N ≤ n ≤ +N und M = 2N + 1.

In the IQC-concept the time operator $T_C$ is given by the matrix

$$C^{km} := \tau^{-1} \sum_{n=-\infty}^{+\infty} \int_{-\tau/2}^{+\tau/2} du \, c^{kn}(u)* \, \tau^n \, c^{mn}(u). \tag{09}$$

The expectation value of $T_C$ at t = 0 is then

$$<\varphi_C(0)|T_C|\varphi_C(0)> = C^{00} = \tau^{-1} \sum_{n=-\infty}^{+\infty} \int_{-\tau/2}^{+\tau/2} du \, c^{0n}(u)* \, \tau^n \, c^{0n}(u) = \tau^{-1} \sum_{n=-\infty}^{+\infty} \tau^n \int_{-\tau/2}^{+\tau/2} du \, |c^{0n}(u)|^2 =$$

$$\sum_{n=-\infty}^{+\infty} \tau^n \frac{1}{2M} \exp(\frac{-|n|}{M}) = 0 \text{ because of } \tau^{-n} = -\tau^n. \tag{10}$$

More generally one gets as a consequence of (06)

$$<\varphi_C(\tau^k)|T_C|\varphi_C(\tau^k)> = C^{kk} = \tau^{-1} \sum_{n=-\infty}^{+\infty} \tau^n \int_{-\tau/2}^{+\tau/2} du \, |c^{kn}(u)|^2 =$$

$$\tau^{-1} \sum_{n=-\infty}^{+\infty} \tau^{(n-k)+k} \int_{-\tau/2}^{+\tau/2} du \, |c^{0n-k}(u)|^2 = \sum_{n=-\infty}^{+\infty} \tau^{(n-k)+k} \frac{1}{2M} \exp(\frac{-|n-k|}{M}) = \tag{11}$$

$$\sum_{n=-\infty}^{+\infty} \tau^{(n-k)} \frac{1}{2M} \exp(\frac{-|n-k|}{M}) + \tau^k \sum_{n=-\infty}^{+\infty} \frac{1}{2M} \exp(\frac{-|n-k|}{M}) = \tau^k.$$

Thus $T_C$ has the correct expectation values in the approximation used here.

## 4. The standard deviation

The variance of $T_C$ in the state $\varphi_C(0)$ is

$$\mathrm{Var}(T_C) = <T_C\varphi_C(0)|T_C\varphi_C(0)> - <\varphi_C(0)|T_C|\varphi_C(0)>^2 = \sum_{n=-\infty}^{+\infty} (C^{0n})*C^{0n} = \qquad (12)$$

$$\sum_{n=-\infty}^{+\infty} |C^{0n}|^2 = \tau^{-2} \sum_{k=-\infty}^{+\infty} \sum_{m=-\infty}^{+\infty} \sum_{n=-\infty}^{+\infty} \int_{-\tau/2}^{+\tau/2} dudv\ c^{0m}(u)*c^{nm}(u)c^{nk}(v)*c^{0k}(v)\ \tau^m\ \tau^k$$

This term from the IQC-concept is too sophisticated for the approach introduced here. Therefore one has to look for an appropriate coarsening of it. The term should be simplified in such a way that only probabilities occur that can be replaced by (05).

This is only possible if $m = n = k$ is set, but the summation over $k$ is maintained. This results in

$$\tau^{-2} \sum_{k=-\infty}^{+\infty} \sum_{n=-\infty}^{+\infty} \int_{-\tau/2}^{+\tau/2} dudv\ c^{0k}(u)*c^{nk}(u)c^{nk}(v)*c^{0k}(v)\ (\tau^k)^2 = \qquad (13)$$

$$\tau^{-2} \sum_{k=-\infty}^{+\infty} (\tau^k)^2 \int_{-\tau/2}^{+\tau/2} du\ c^{0k}(u)*c^{00}(u) \int_{-\tau/2}^{+\tau/2} dv\ c^{00}(v)*c^{0k}(v),$$

because of $c^{kk}(u) = c^{00}(u)$, $c^{kk}(v) = c^{00}(v)$.

Further it is assumed that values $w_k \in [-\tau/2; +\tau/2]$ do exist so that

$$\tau^{-1} \int_{-\tau/2}^{+\tau/2} du \, c^{0k}(u)*c^{00}(u) = c^{0k}(w_k)*c^{00}(w_k), \text{ and}$$

$$\tau^{-1} \int_{-\tau/2}^{+\tau/2} dv \, c^{00}(v)*c^{0k}(v) = c^{00}(w_k)*c^{0k}(w_k). \text{ This yields}$$

$$\sum_{k=-\infty}^{+\infty} (\tau^k)^2 \, c^{0k}(w_k)*c^{00}(w_k) \, c^{00}(w_k)*c^{0k}(w_k) = \sum_{k=-\infty}^{+\infty} (\tau^k)^2 \, |c^{00}(w_k)|^2 |c^{0k}(w_k)|^2. \tag{14}$$

With $|c^{0k}(w_k)|^2 = \frac{1}{2M} \exp(\frac{-|k|}{M})$ and the corresponding $|c^{00}(w_k)|^2 = \frac{1}{2M}$ follows

(the special choice of $w_k \in [-\tau/2; +\tau/2]$ does not matter for this approximation)

$$\text{Var}(T_C) = (\frac{1}{2M})^2 \sum_{k=-\infty}^{+\infty} (\tau^k)^2 \exp(\frac{-|k|}{M}) = (\tau_0^2/4M^4) \sum_{k=-\infty}^{+\infty} k^2 \exp(\frac{-|k|}{M}). \tag{15}$$

The sum is the second derivative of a geometric series and yields

$$\text{Var}(T_C) = (\tau_0^2/4M^4) \, [\exp(M^{-1}) + \exp(2M^{-1})][\exp(M^{-1}) - 1]^{-3} \rightarrow \tau_0^2/M \tag{16}$$

for very large M so that the standard deviation is

$$\sigma(T_C) = \tau_0/\sqrt{M} = \tau\sqrt{M}. \tag{17}$$

## 5. Accuracy limits of time measurement.

The minimalization of the standard deviation (17) requires a small time $\tau_0 := 1/\lambda$ of the elementary processes used and a huge number M of them. In [2, 3] is pointed out that M is limited from physical reasons. The time span $\tau := \tau_0/M$ must far exceed the Planck time. All this results in a lower limit for the standard deviation. However, because here a general group of quantum clocks is considered, the question arises, whether or not the same lower limit applies to the standard deviation of all such stochastic systems.

So while the standard deviation cannot be reduced arbitrarily, a possible time dilatation accordingly affects the standard deviation. Any time dilation, whether caused by gravitational fields [4] or extreme velocities of the device, causes an increasing uncertainty in time measurement by the quantum clock. This may affect satellite time measurements in space navigation. A time measurement near the Schwarzschild spherical shell of a black hole becomes completely blurred and thus impossible.